\documentclass[%
preprint,
amsmath,amssymb,
aps,
pra,
floatfix,
]{revtex4-2}

\usepackage{graphicx}
\usepackage{dcolumn}
\usepackage{bm}
\usepackage{textgreek}

\begin{document}

\preprint{}

\title{Cartesian and spherical multipole expansions\\in anisotropic media}

\author{Elias Le Boudec}
\email{elias.leboudec@alumni.epfl.ch}
\author{Toma Oregel-Chaumont}
\author{Farhad Rachidi}
\affiliation{EPFL, 1015 Lausanne, Switzerland}
\author{Marcos Rubinstein}
\affiliation{University of Applied Sciences and Arts Western Switzerland, 1401 Yverdon-les-Bains, Switzerland}
\author{Felix Vega}
\affiliation{Technology Innovation Institute, Abu Dhabi, United Arab Emirates} 
\date{\today}
\begin{abstract} 
The multipole expansion can be formulated in spherical and Cartesian coordinates. By constructing an explicit map linking both formulations in isotropic media, we discover a lack of equivalence between them in anisotropic media. 
In isotropic media, the Cartesian multipole tensor can be reduced to a spherical tensor containing fewer independent components.
In anisotropic media, however, the loss of propagation symmetry prevents this reduction. Consequently, non-harmonic sources radiate fields that can be projected onto a finite set of Cartesian multipole moments but require potentially infinitely many spherical moments. For harmonic sources, the link between the two approaches provides a systematic way to construct the spherical multipole expansion from the Cartesian one.
The lack of equivalence between both approaches results in physically significant effects wherever the field propagation includes the Laplace operator.  We demonstrate this issue in an electromagnetic radiation inverse problem in anisotropic media, including an analysis of a large-anisotropy regime and an introduction to vector spherical harmonics. We show that the use of the Cartesian approach increases the efficiency and interpretability of the model. The proposed approach opens the door to a broader application of the multipole expansion in anisotropic media.
\end{abstract}

\maketitle



\section{Introduction}
\label{sec: introduction}

The multipole expansion method has long been an essential mathematical tool successfully applied to a wide range of systems described by partial differential equations involving the Laplace operator. 
Commonly used in electromagnetism~\cite{french_theory_1953,wallace_theory_1951}, this method also sees applications from the quantum level~\cite{griffiths_introduction_1995} up to Newtonian gravitational interaction~\cite{stirling_multipole_2017,stirling_closed_2019} and gravitational waves~\cite{thorne_multipole_1980,zhang_multipole_1986}; indeed, whenever one needs to describe how fields behave outside the source region.
However, the multipole expansion in anisotropic media, where wave propagation becomes direction-dependent, has remained relatively ill-studied. 
This must be distinguished from the multipole expansion associated with an anisotropic source, which results in direction-dependent radiation or scattering \cite{evlyukhin_multipole_2019}. 
Here, we assume that wave propagation occurs within an anisotropic medium, while the source itself may also exhibit anisotropic characteristics.

For optical or radio-frequency systems, the rise of advanced wave manipulation, such as via metamaterials~\cite{smith_metamaterials_2004,zheludev_metamaterials_2012,terekhov_multipole_2019} or nanophotonics~\cite{alaee_exact_2019}, warrants a study of this method in anisotropic media. For instance, meta-atoms mixing electric and magnetic multipole moments give rise to bianisotropy \cite{poleva_multipolar_2023}. 
The multipole expansion is also readily applied to metasurfaces to express the fields, which are then linked through the $T$-matrix method~\cite{rahimzadegan_comprehensive_2022}. 
For example, this approach has been applied to the analysis of substrate-supported dielectric nanoresonator metasurfaces~\cite{czajkowski_multipole_2020}. The multipole expansion is also used to obtain a basis of resonant states in the resonant-state expansion method for the analysis of open optical systems~\cite{doost_resonant-state_2014}. 
High-order multipoles are also gaining interest in connection to the chosen center of coordinates \cite{ospanova_modified_2023}.
A variation of the multipole expansion can also be used to describe the nonlocal response of certain metasurfaces~\cite{achouri_spatial_2023}. While the literature commonly uses the multipole expansion to describe and analyze metamaterials with anisotropic wave propagation, the analysis of the multipole expansion of fields \emph{embedded} in an anisotropic medium is missing. Implications of this analysis could include the effect of a layered substrate with uniaxial wave propagation on the interaction between neighboring particles in a metasurface.

The analogous nature of electromagnetic and sound waves means that anisotropic wave propagation can also occur within acoustic metamaterials~\cite{smith_tailored_2022}.
In plasma physics as well, the multipole expansion is used to describe the interaction potentials of dust particles~\cite{ramazanov_multipole_2016}, though the method has yet to see application to anisotropic cases beyond the first order~\cite{lapenta_dipole_1995}.
On the largest scales, anisotropic baryon acoustic oscillations imprint on the clustering of galaxies and can be observed in their power spectra~\cite{taruya_forecasting_2011},
while the cosmic neutrino background applies an anisotropic stress that dampens gravitational waves, a potentially observable effect in the multipole expansion of the cosmic microwave background~\cite{bernardini_cb_2014}.

In the above-mentioned studies, solutions of wave-like equations featuring the Laplace operator are found using the separation of variables technique and representing the field as a linear combination of harmonic polynomials (for the angular part) and spherical Hankel functions of the first kind, second kind, or a combination of both (for the radial part)~\cite{jackson_classical_1999}, yielding the spherical multipole expansion. 
Equivalently, the same solutions can be found by using a Taylor series expansion of Green's function (see, e.g.,~\cite{kocher_point-multipole_1978}), yielding the Cartesian multipole expansion. 
The number of Cartesian basis functions (and moments) grows cubically with the series truncation order, while the same growth is only quadratic for the spherical expansion. The equivalence of both for electrostatics is the subject of Problem~4.3 in~\cite{jackson_classical_1999}. 
Indeed, the solution of this exercise shows that any Cartesian moments tensor of order $\ell $ can be represented by a linear combination of spherical moments tensors of order $\ell $, $\ell -2$, etc., down to 0 or 1. 
For plasma physics and the Vlasov–Fokker–Planck equation, a detailed computation of the conversion between both approaches is presented in~\cite{schween_converting_2022}; the references therein, e.g.,~\cite{johnston_cartesian_1960}, are also of interest. Contrary to~\cite{schween_converting_2022}, we are interested in the singular behavior of multipole expansions, i.e., their tendency to explode at the origin. 
This allows us to consider all Cartesian moments, including, for example, the trace of the second-order moments tensor, which is set to zero by design in other references. 
While this is necessary for isotropic media to satisfy the wave propagation spherical symmetry, we need to relax this assumption in anisotropic media to obtain a full representation of fields in a general setting.
 Note that the presented limitation of the spherical formulation is independent of the (non)existence of analytical closed-form expressions of Green's functions; rather, it is a consequence of the loss of symmetry in the wave propagation. 
 In isotropic media, the spherical symmetry of the Laplace operator implies that irreducible tensors are sufficient to describe scattered or radiated fields. 
 As we will see, in anisotropic media, all tensor elements are necessary to describe the fields fully. 
 As such, we show that the Cartesian multipole expansion is central to obtaining field solutions in anisotropic media.
 
In this paper, we describe the link between the spherical and Cartesian multipole expansions in isotropic media and demonstrate this equivalence breaks down in anisotropic media. 
 Next, we also discuss the link to the vector spherical harmonics: the extension of their scalar counterpart to vector fields. We then  illustrate the lack of equivalence in a radiation inverse problem, including a detailed analysis of a large-anisotropy regime.
 We show that, in anisotropic media, the use of the Cartesian multipole expansion improves the modeling efficiency and interpretability.

\section{Methods}
\label{sec: derivation}
\subsection{Spherical and Cartesian multipole expansions in isotropic media}
We begin by finding equivalence relations between the spherical and Cartesian multipole expansions by analyzing their singular behavior. This will allow us to extrapolate the approach to anisotropic media by considering the well-defined radiation of point sources. To this end, previous work on the pseudopotential method~\cite{derevianko_revised_2005,idziaszek_pseudopotential_2006,ghosh_roy_pseudopotential_2013} has sparked interest in the singular behavior of spherical wave functions. This behavior is challenging to describe as the concept of ``radial'' derivatives of the Dirac $\delta$ function is not trivially defined (see~\cite{yang_distributions_2013, brackx_radial_2017, estrada_regularization_2017}) because of the singular nature of spherical coordinates at the origin. A rigorous approach in
\cite{stampfer_mathematically_2008} allows to write, for the spherical basis function $f_{\ell m}(\mathbf{r})$ introduced in Equation~(\ref{eq: def of spherical solution}),
\begin{equation}\label{eq: singular behavior spherical}
    (\nabla^2+k_0^2)f_{\ell m}(\mathbf{r})=\frac{(-1)^\ell  4\pi \text{i}}{k_0^{\ell +1}}Y_{\ell m}(\nabla)\delta(\mathbf{r})=:\tilde{Y}_{\ell m}(\nabla)\delta(\mathbf{r})
\end{equation}
where $\nabla$ is the gradient operator, $k_0$ is the free-space wavenumber, $\delta(\mathbf{r})$ is the Dirac $\delta$-function (a distribution), and the solid harmonics $Y_{\ell m}(\mathbf{k})=k^\ell  \mathcal{Y}_{\ell m}(\theta,\phi)$ are the homogeneous and harmonic polynomials obtained from the spherical harmonics $\mathcal{Y}_{\ell m}(\theta,\phi)$. $(k,\theta,\phi)$ are the spherical coordinates of the  three-dimensional vector $\mathbf{k}$. We use the convention
\begin{equation}
    \mathcal{Y}_{\ell m}(\theta,\phi)=\sqrt{\frac{2\ell+1}{4\pi}}\sqrt{\frac{(l-m)!}{(l+m)!}}P_{\ell m}\left[\cos(\theta)\right]\text{e}^{\text{i}m\phi}
\end{equation}
where $P_{\ell m}(z)$ is the associated Legendre polynomial.
$Y_{\ell m}(\nabla)$ is a partial differential operator obtained by replacing the Cartesian components of $\mathbf{k}=(k_x,k_y,k_z)$ in $Y_{\ell m}(\mathbf{k})$ by
\begin{equation}
    k_x\to\partial/\partial x,~k_y\to\partial/\partial y,~k_z\to\partial/\partial z
\end{equation}
and likewise for $\tilde{Y}_{\ell m}(\nabla)$. For instance, $Y_{1,1}(\mathbf{k})=-1/2\sqrt{3/(2\pi)}(k_x+\text{i}k_y)$, hence $Y_{1,1}(\nabla)\delta(\mathbf{r})=-1/2\sqrt{3/(2\pi)}(\partial\delta(\mathbf{r})/\partial x+\text{i}\partial\delta(\mathbf{r})/\partial y)$.

To obtain the singular behavior in Equation~(\ref{eq: singular behavior spherical}), we start by noticing that the function $r^{-1}$ yields a Dirac singularity under the Laplace operator:
\begin{equation}
    \nabla^2\text{vp}(r^{-1})=-4\pi \delta(\mathbf{r})
\end{equation}
$\text{vp}$ denotes the principal value. This generalizes to higher orders as \cite[Lemma~1.(c)]{stampfer_mathematically_2008}
\begin{equation}
    \nabla^2 \text{vp}\left[r^{-\ell-1}\mathcal{Y}_{\ell m}(\theta,\phi)\right]=\frac{(-1)^{\ell-1}4\pi}{(2\ell-1)!!}{Y}_{\ell m}(\nabla)\delta(\mathbf{r})
\end{equation}
where $(2\ell-1)!!=1\cdot3\cdot5\cdots(2\ell-1)$. Moving to waves and the Helmholtz operator, the inverse radius is replaced with the spherical Neumann function $y_\ell(k_0r)$ which also behaves in $(k_0r)^{-\ell-1}$ as $r\to0$. A consequence of the same Lemma is that
\begin{equation}
    (\nabla^2+k_0^2)\text{vp}\left[y_\ell(k_0r)\mathcal{Y}_{\ell m}(\theta,\phi)\right]=\frac{(-1)^{\ell}4\pi}{k_0^{\ell+1}}{Y}_{\ell m}(\nabla)\delta(\mathbf{r})
\end{equation}
Radiating waves are described by spherical Hankel functions of the first or second kind $h_\ell^{(1,2)}(k_0r)$ \cite[Eq.~10.1.1]{abramowitz_handbook_1972}:
\begin{equation}
    h_\ell^{(1)}(k_0r)=j_\ell(k_0r)+\text{i}y_\ell(k_0r)
\end{equation}
where $j_\ell(k_0r)$ is a spherical Bessel function of the first kind. As this function is regular at the origin, the complete radiating wave is
\begin{equation}\label{eq: def of spherical solution}
    f_{\ell m}(\mathbf{r}):=\text{vp}\left[h_\ell^{(1)}(k_0r)\mathcal{Y}_{\ell m}(\theta,\phi)\right]    
\end{equation}
and its singular behavior is given by Equation~(\ref{eq: singular behavior spherical}).

On the other hand, the singular behavior of the Cartesian basis functions is simply given, by definition, by (see, e.g.,~\cite{wunsche_schwache_1975,le_boudec_time-domain_2024})
\begin{equation}\label{eq: def of cartesian solution}
    (\nabla^2+k_0^2)f_{\alpha}(\mathbf{r})=\nabla^\alpha\delta(\mathbf{r})
\end{equation}
where we have used the multi-index notation, i.e., $\alpha=(\alpha_x,\alpha_y,\alpha_z)\in\mathbb{N}^3$, and $\nabla^\alpha=(\partial/\partial x)^{\alpha_x}(\partial/\partial y)^{\alpha_y}(\partial/\partial z)^{\alpha_z}$. The Cartesian solutions $f_\alpha(\mathbf{r})$ can be obtained directly from Green's function $G(\mathbf{r})$, which satisfies by definition
\begin{equation}
    (\nabla^2+k_0^2)G(\mathbf{r})=\delta(\mathbf{r})
\end{equation}
Indeed, since
\begin{equation}
    \nabla^\alpha\left[(\nabla^2+k_0^2)G(\mathbf{r})\right]=(\nabla^2+k_0^2)\nabla^\alpha G(\mathbf{r})=\nabla^\alpha\delta(\mathbf{r})
\end{equation}
we have
\begin{equation}
    f_\alpha(\mathbf{r})=\nabla^\alpha G(\mathbf{r}).
\end{equation}

The next step is to assume that under appropriate boundary conditions (e.g., radiation conditions for the spherical Hankel function of the first or second kind), there are unique solutions  $f_{\ell m}(\mathbf{r})$ and $f_\alpha(\mathbf{r})$ to the wave equation with sources, equations~(\ref{eq: singular behavior spherical})~and~(\ref{eq: def of cartesian solution}). Conversely, the Helmholtz operator applied to the solutions $f_{\ell m}(\mathbf{r})$ and $f_\alpha(\mathbf{r})$ yields singularities that are entirely determined by the polynomials $\tilde{Y}_{\ell m}(\mathbf{k})$ and $\mathbf{k}^\alpha$. Hence, from now on, we identify spherical solutions $f_{\ell m}(\mathbf{r})$ by their corresponding harmonic polynomial $\tilde{Y}_{\ell m}(\mathbf{k})$ and the Cartesian solutions $f_\alpha(\mathbf{r})$ by the monomials $\mathbf{k}^\alpha$:
\begin{align}\label{eq: flm and y(k)}
    f_{\ell m}(\mathbf{r})=\text{vp}\left[h_\ell^{(1)}(k_0r)\mathcal{Y}_{\ell m}(\theta,\phi)\right]&\iff \tilde{Y}_{\ell m}(\mathbf{k})\\
    f_\alpha(\mathbf{r})=\nabla^\alpha G(\mathbf{r})&\iff \mathbf{k}^\alpha \label{eq: falpha and kalpha}
\end{align}

With  this formalism in mind, we now strive to write equivalences between the spherical and Cartesian solutions by finding equivalences between their corresponding polynomials. 

\subsubsection{Spherical to Cartesian multipole expansions}
Clearly, any harmonic polynomial can be written as a linear combination of monomials; hence, any spherical solution can be represented as a linear combination of Cartesian solutions. For example, since \begin{equation}
\tilde{Y}_{2,0}(\mathbf{k})=\text{i}\sqrt{5\pi}/k_0^3(2k_z^2-k_x^2-k_y^2)
\end{equation}
we have
\begin{equation}\label{eq: ex spherical to Cartesian}
    f_{2,0}(\mathbf{r})=\text{i}\sqrt{5\pi}/k_0^3\left[2f_{(0,0,2)}(\mathbf{r})-f_{(2,0,0)}(\mathbf{r})-f_{(0,2,0)}(\mathbf{r})\right].
\end{equation}

\subsubsection{Cartesian to spherical multipole expansions}
Nevertheless, it is generally not possible to write an arbitrary monomial $\mathbf{k}^\alpha$ as a linear combination of harmonic polynomials $\tilde{Y}_{\ell m}(\mathbf{k})$, as the monomial is not necessarily harmonic. For any monomial,
we can nonetheless write  by \cite[Theorem~5.21]{axler_harmonic_2013}:
\begin{equation}\label{eq: k^alpha as sum of harmonic polynomials}
    \mathbf{k}^\alpha=\sum_{n=0}^{\left\lfloor|\alpha|/2\right\rfloor}(k^2)^n p_{|\alpha|-2n}(\mathbf{k})
\end{equation}
where the harmonic polynomials $p_{|\alpha|-2n}(\mathbf{k})$ are given by
\begin{equation}
    p_{|\alpha|-2n}(\mathbf{k})=\sum_{j=n}^{\left\lfloor|\alpha|/2\right\rfloor}c_{j,n}k^{2(j-n)}({\nabla_\mathbf{k}}^2)^j\mathbf{k}^\alpha
\end{equation}
 $|\alpha|=\alpha_x+\alpha_y+\alpha_z$, $\lfloor\cdot\rfloor$ denotes the floor function, ${\nabla_\mathbf{k}}^2$ is the Laplace operator with respect to $\mathbf{k}$, and $c_{j,n}$ are coefficients specified in the cited theorem, which are omitted here for brevity.

Next, we project the harmonic polynomials $p_{|\alpha|-2n}(\mathbf{k})$ onto the solid harmonics ${Y}_{|\alpha|-2n,m}(\mathbf{r})$
\begin{equation}\label{eq: p as solid harmonics}
    p_{|\alpha|-2n}(\mathbf{k})=\sum_{m=-|\alpha|+2n}^{|\alpha|-2n}A_{|\alpha|-2n,m}Y_{|\alpha|-2n,m}(\mathbf{k})
\end{equation}
by using the orthonormality of the spherical harmonics ${\mathcal{Y}}_{\ell,m}(\theta,\phi)$:
\begin{equation}
    A_{|\alpha|-2n,m}=\int_0^\pi\text{d}\theta\int_0^{2\pi}\text{d}\phi\sin(\theta)p_{|\alpha|-2n}(\mathbf{k})/k^{|\alpha|-2n}\mathcal{Y}^*_{|\alpha|-2n,m}(\theta,\phi)
\end{equation}
where $\cdot^*$ denotes complex conjugation, and $p_{|\alpha|-2n}(\mathbf{k})/k^{|\alpha|-2n}$ is the angular part of the polynomial $p_{|\alpha|-2n}(\mathbf{k})$ (a function of $\theta$ and $\phi$). Combining equations~(\ref{eq: k^alpha as sum of harmonic polynomials}) and~(\ref{eq: p as solid harmonics}), we find
\begin{equation}
    \mathbf{k}^\alpha=\sum_{n=0}^{\left\lfloor|\alpha|/2\right\rfloor}(k^2)^n \sum_{m=-|\alpha|+2n}^{|\alpha|-2n}A_{|\alpha|-2n,m}Y_{|\alpha|-2n,m}(\mathbf{k})
\end{equation}
Applying the scaling defined in Equation~(\ref{eq: singular behavior spherical}), we have
\begin{equation}
    \mathbf{k}^\alpha=\sum_{n=0}^{\left\lfloor|\alpha|/2\right\rfloor}(k^2)^n \sum_{m=-|\alpha|+2n}^{|\alpha|-2n}\tilde{A}_{|\alpha|-2n,m}\tilde{Y}_{|\alpha|-2n,m}(\mathbf{k})
\end{equation}
where
\begin{equation}
    \tilde{A}_{|\alpha|-2n,m}=\frac{k_0^{|\alpha|-2n+1}}{(-1)^{|\alpha|-2n} 4\pi \text{i}}A_{|\alpha|-2n,m}
\end{equation}

 Now, by using the equivalence between the polynomials and their corresponding multipole expansion (see equations~(\ref{eq: flm and y(k)}) and~(\ref{eq: falpha and kalpha})), we can write any Cartesian solution as
\begin{equation}\label{eq: f alpha as lin comb of f lm}
f_\alpha(\mathbf{r})=\sum_{n=0}^{\left\lfloor|\alpha|/2\right\rfloor} (\nabla^{2})^n\sum_{m=-|\alpha|+2n}^{|\alpha|-2n}\tilde{A}_{|\alpha|-2n,m}f_{|\alpha|-2n,m}(\mathbf{r})
\end{equation}
Next, for wave-like equations, like the Helmholtz equation, the Laplacian $\nabla^2$ of solutions can be re-written from Equation~(\ref{eq: def of spherical solution}):
\begin{equation}\label{eq: laplacian of spherical sol}
    \nabla^2f_{\ell m}(\mathbf{r})=-k_0^2f_{\ell m}(\mathbf{r})+\tilde{Y}_{\ell m}(\nabla)\delta(\mathbf{r})
\end{equation}
Ignoring the origin, we have
\begin{equation}\label{eq: laplacian of spherical sol no origin}
    \nabla^2f_{\ell m}(\mathbf{r})=-k_0^2f_{\ell m}(\mathbf{r})
\end{equation}
Inserting into Equation~(\ref{eq: f alpha as lin comb of f lm}), we obtain
\begin{equation}\label{eq: spherical to cartesian full outside origin}
f_\alpha(\mathbf{r})=\sum_{n=0}^{\left\lfloor|\alpha|/2\right\rfloor}\sum_{m=-|\alpha|+2n}^{|\alpha|-2n}\tilde{A}_{|\alpha|-2n,m}(-k_0^2)^nf_{|\alpha|-2n,m}(\mathbf{r})
\end{equation}
outside the origin. In other words, any Cartesian solution of order $|\alpha|=\ell $ can be written as a linear combination of spherical solutions of order $\ell $, $\ell -2$, etc., down to $\ell =0$ or $\ell =1$. For example, the non-harmonic octupole $\alpha=(0,0,2)$ (identified by the monomial $\mathbf{k}^\alpha=k_z^2$) is projected onto solid spherical harmonics of orders zero and two as
\begin{equation}
k_z^2=2/3\left[\sqrt{\pi}k^2 Y_{0,0}(\mathbf{k})+2\sqrt{\pi/5}Y_{2,0}(\mathbf{k})\right]    
\end{equation}
which corresponds to 
\begin{equation}
k_z^2=-\text{i} k_0/6\left[1/\sqrt{\pi}k^2 \tilde{Y}_{0,0}(\mathbf{k})+2k_0^2/\sqrt{5\pi}\tilde{Y}_{2,0}(\mathbf{k})\right]
\end{equation}
By Equation~(\ref{eq: laplacian of spherical sol no origin}), multiplying the polynomial $\tilde{Y}_{0,0}(\mathbf{k})$ by $k^2$ is equivalent to multiplying the solution $f_{0,0}(\mathbf{r})$ by $-k_0^2$, and we obtain
\begin{equation}
f_{(0,0,2)}(\mathbf{r})=\text{i} k_0^3/6\left[1/\sqrt{\pi} f_{0,0}(\mathbf{r})-2/\sqrt{5\pi}f_{2,0}(\mathbf{r})\right]
\end{equation}
outside the origin. At the origin, derivatives of the Dirac $\delta$ function must also be added in compliance with Equation~(\ref{eq: laplacian of spherical sol}).

\subsection{The case of anisotropic media}
In anisotropic media, the Laplace operator (as in Equation~(\ref{eq: laplacian of spherical sol})) is replaced with weighted second-order derivatives, reflecting different propagation characteristics along different directions. For example, electromagnetic fields obey a modified vector wave equation \cite[Equation~(5.6)] {mackay_electromagnetic_2019} whose anisotropy is determined by the constitutive parameters' tensorial nature. In such cases, as we illustrate in the section ``Illustration on an inverse problem,'' some sources radiate fields that cannot be fully represented by a spherical multipole expansion. In turn, the more numerous Cartesian moments are necessary to model these fields accurately and efficiently.

\subsection{Relation to the vector spherical harmonics}
Many applications such as optics require the use of vector spherical harmonics. Here, we wish to outline how the link between the spherical and Cartesian approaches can be extended to the vector spherical harmonics and, thus, the vector wave equation. 

\subsubsection{From vector spherical harmonics to the Cartesian multipole expansion}
Fortunately, it is possible to write the components of vector spherical harmonics in terms of scalar spherical harmonics. This allows to generalize the reasoning applied to the scalar wave equation. Using quantum theory formalism and the Condon–Shortley phase, the vector spherical harmonics can be written as \cite[Eq.~7.3.3]{varshalovich_quantum_1988}
\begin{equation}\label{eq: vector spherical harmonics as scalar spherical harmonics}
    \bm{\mathcal{Y}}^\ell_{j m}(\theta,\phi)=\sum_{m',q}C^{jm}_{\ell m' 1q}\mathcal{Y}_{\ell m'}(\theta,\phi)\mathbf{e}_q
\end{equation}
where $j=\ell,\ell\pm1$ is the total angular momentum quantum number, $\ell$ the orbital angular momentum quantum number, $m=-j,\dots,j$, the sum runs over $m'=-\ell,\dots\ell$, $q=\pm1,0$, $C^{jm}_{\ell m 1q}$ denotes the Clebsch-Gordan coefficient, and the $\mathbf{e}_q$ form the covariant spherical basis defined in Cartesian coordinates as
\begin{align}
    &\mathbf{e}_{\pm1}=\mp\frac{1}{\sqrt{2}}\left(\mathbf{e}_x\pm \text{i}\mathbf{e}_y\right)\\
    &\mathbf{e}_0=\mathbf{e}_z
\end{align}
The vector spherical harmonics used by Jackson \cite{jackson_classical_1999} $\bm{\mathcal{Y}}_{\ell m}^{(q)}(\theta,\phi)$, $q=0,\pm1$, are obtained as linear combinations of   $\bm{\mathcal{Y}}^\ell_{j m}(\theta,\phi)$ (see supplementary equations~(S1-S3)). They  correspond to the azimuthal ($q=0$), polar ($q=1$), and radial ($q=-1$) components. In the context of Mie scattering, $\bm{\mathcal{Y}}_{\ell m}^{(1)}(\theta,\phi)$ represents electric multipoles, whereas $\bm{\mathcal{Y}}_{\ell m}^{(0)}(\theta,\phi)$ corresponds to magnetic multipoles. In homogeneous media, the radial component is not allowed for electromagnetic fields. Together with Equation~(\ref{eq: vector spherical harmonics as scalar spherical harmonics}), we show in the Supplementary Note~1 how to express the azimuthal or polar vector spherical harmonics using the Cartesian multipole expansion of an appropriate current source. We show that a current density given by
\begin{equation}
   \mathbf{J}_{\ell m}^{(0)}(\mathbf{r})\propto\mathbf{Y}_{\ell m}^{\ell}(\nabla)\delta(\mathbf{r})
\end{equation}
radiates an electric field whose angular dependence is given by the azimuthal vector spherical harmonic $\bm{\mathcal{Y}}_{\ell m}^{(0)}(\theta,\phi)$. The solid vector harmonics correspond to  $\mathbf{Y}_{\ell m}^{\ell}(\mathbf{k})=k^\ell\bm{\mathcal{Y}}_{\ell m}^\ell(\mathbf{k})$. On the other hand, the polar vector spherical harmonic is radiated by 
\begin{equation}
   \mathbf{J}_{\ell m}^{(1)}(\mathbf{r})\propto\mathbf{Y}_{\ell\pm1, m}^{\ell}(\nabla)\delta(\mathbf{r})
\end{equation}
where $\mathbf{Y}_{\ell \pm1,m}^{\ell}(\mathbf{k})=k^{\ell}\bm{\mathcal{Y}}_{\ell\pm1, m}^{\ell}(\mathbf{k})$. The Cartesian components of the current densities $\mathbf{J}_{\ell m}^{(0,1)}(\mathbf{r})$ correspond to singularities of the scalar wave equation analyzed in the ``Spherical and Cartesian multipole expansions in isotropic media'' subsection of the ``Methods'' section. Hence, any component of the vector spherical harmonics can be represented by an appropriate sum of Cartesian multipoles.

\subsubsection{From the Cartesian multipole expansion to the vector spherical harmonics}
As expected from the findings for the scalar wave equation, the inverse transformation---i.e., writing an arbitrary vector Cartesian multipole expansion as a linear combination of vector spherical harmonics---is not as straightforward, and a complete treatment goes beyond the scope of this research. To understand the issue at hand, one approach is to count the number of vector spherical harmonics at a given order and to compare with the corresponding number of Cartesian multipole moments. As we will see, as in the scalar case, there is an excess of Cartesian multipole moments. 

Let us now count the vector spherical harmonics. The azimuthal vector harmonics correspond to a current density given by the vector spherical harmonics $\bm{\mathcal{Y}}_{\ell m}^{\ell}(\theta,\phi)$. In light of Equation~(\ref{eq: vector spherical harmonics as scalar spherical harmonics}), there are $2\ell+1$ different azimuthal waves of order $\ell>0$ (none for $\ell=0$). By the same argument, there are $2(\ell\pm1)+1$ polar vector harmonics of order $\ell>0$ corresponding to $j=\ell\pm1$. But the polar harmonics are degenerate, in that both values of $j$ yield the same waves, by virtue of the contraction in Equation~(\ref{eq: laplacian of spherical sol no origin}) (see the Supplementary Note~1.2.3). Thus, in total, there are
\begin{equation}
    [2\ell+1]+[2(\ell+1)+1]=4\ell+4
\end{equation}
vector spherical harmonics of order $\ell>0$. For $\ell=0$, only the polar solutions exist, and their count is given by $2\cdot(0+1)+1=3$.

Next, we count the moments for the vector Cartesian multipole expansion. Since there are three possible polarizations ($x$, $y$, and $z$) and $2\ell+1$ spherical harmonics of order $\ell$ per polarization, one would expect
\begin{equation}
    3(2\ell+1)=6\ell+3
\end{equation}
different vector spherical harmonics (VSH). The missing harmonics are the degenerate polar harmonics with $j=\ell-1$, and their count is given by $2(\ell-1)+1$:
\begin{equation}
    \underbrace{4\ell+4}_{\text{VSH}}+\underbrace{2(\ell-1)+1}_\text{degenerate}=\underbrace{6\ell+3}_\text{Cartesian}
\end{equation}
This hints towards a similar issue to the scalar case when going to anisotropic media, as the Laplace operator is responsible for the degeneracy. Another viewpoint is that in anisotropic media, the degeneracy disappears, and new degrees of freedom in wave polarization become available.

To note, we circumvent this issue in the section ``Illustration on an inverse problem'' by opting for the electric Green's tensor-valued function instead of vector spherical harmonics. This approach enables a direct connection between the electric current and the vector-valued field.
\section{Illustration on an inverse problem}
\label{sec: Illustration}

Here, we show that the loss of equivalence between spherical and Cartesian multipole expansions has measurable and significant effects on an electromagnetic radiation inverse problem. We consider an electric uniaxial medium characterized by the relative permittivity tensor
\begin{equation}
\overline{\overline{\varepsilon}}=\mathbf{e}_x\mathbf{e}_x^\top+\mathbf{e}_y\mathbf{e}_y^\top+\varepsilon_{zz}\mathbf{e}_z\mathbf{e}_z^\top
\end{equation}
On top of appearing in natural materials (e.g., calcite~\cite{hotchkiss_table_1908}), electromagnetic anisotropic media commonly describe metamaterials (e.g., through effective medium theory~\cite{jiang_effective_2022}), which are seeing an increase of adoption from optical~\cite{koepfli_metamaterial_2023} to radio frequencies~\cite{khan_eight-port_2023}. High anisotropy can be achieved through 3D printing of carbon fiber composite filament \cite{harmon_permittivity_2023}. 

To illustrate, we consider a radiation inverse problem---i.e., we try to predict simulation results (the ground-truth data) by performing a multipole expansion in spherical and Cartesian coordinates. For $\varepsilon_{zz}\leq 10$, the ground truth data consists of the electric field on a sphere surrounding the source
\begin{equation}
    \mathbf{J}(\mathbf{r})=\nabla^\alpha\delta(\mathbf{r})\mathbf{e}_x
\end{equation}
at a distance of one vacuum wavelength $\lambda_0$ simulated using Meep~\cite{oskooi_meep_2010} in the frequency domain. The simulation domain consists of a cube of side $2\lambda_0$ surrounded by a perfectly matched layer of thickness $\lambda_0/5$.
The source consists, first, of a pure dipole ($\alpha=(0,0,0)$) and then of a non-harmonic octupole ($\alpha=(0,0,2)$). The latter is approximated by four dipole sources of alternating signs separated by a simulation voxel along the $z$ axis.

To avoid the ``inverse crime,''~\cite{colton_inverse_2019} the fitting data (i.e., the prediction through a multipole expansion) is obtained from the analytical Green's function~\cite{chen_theory_1983}. The Cartesian multipole expansion basis functions are obtained by differentiating Green's function up to a fixed order $\ell _\text{max}$, i.e.,
\begin{equation}
    \mathbf{f}_\alpha(\mathbf{r})=\nabla^\alpha \mathbf{G}(\mathbf{r})\mathbf{e}_x
\end{equation}where $\mathbf{G}(\mathbf{r})$ is Green's tensor-valued function. $\mathbf{G}$ is defined as the solution of
\begin{equation}
    (\nabla\nabla^\top-\nabla^2\mathbf{I}-k_0^2\overline{\overline{\varepsilon}})\mathbf{G}(\mathbf{r})=\mathbf{I}\delta(\mathbf{r)}
\end{equation}
where $\mathbf{I}$ is the identity matrix. In turn,
\begin{equation}
    \mathbf{E}(\mathbf{r})=-\text{i}\omega\mu\iiint\text{d}^3\mathbf{r}'\mathbf{G}(\mathbf{r}-\mathbf{r}')\mathbf{J}(\mathbf{r}')
\end{equation}
and $\mu$ is the medium's magnetic permeability. We use the following convention for the Fourier transform $f(\omega)$ of a function $f(t)$:
\begin{equation}
    f(\omega)=\int \text{d}tf(t)\text{e}^{-\text{i}\omega t}
\end{equation}
For $\varepsilon_{zz}>10$, the ground truth data is also obtained from the analytical Green's function. Doing so, while the ``inverse crime'' is committed, the data range can be extended to high degrees of anisotropy (up to $\varepsilon_{zz}=10^3$) while avoiding prohibitively large numerical problems. 

In parallel, the spherical multipole expansion basis functions are derived as linear combinations of the Cartesian basis functions, as in Equation~(\ref{eq: ex spherical to Cartesian}). 
\begin{figure*}[p]
\includegraphics[scale=.8]{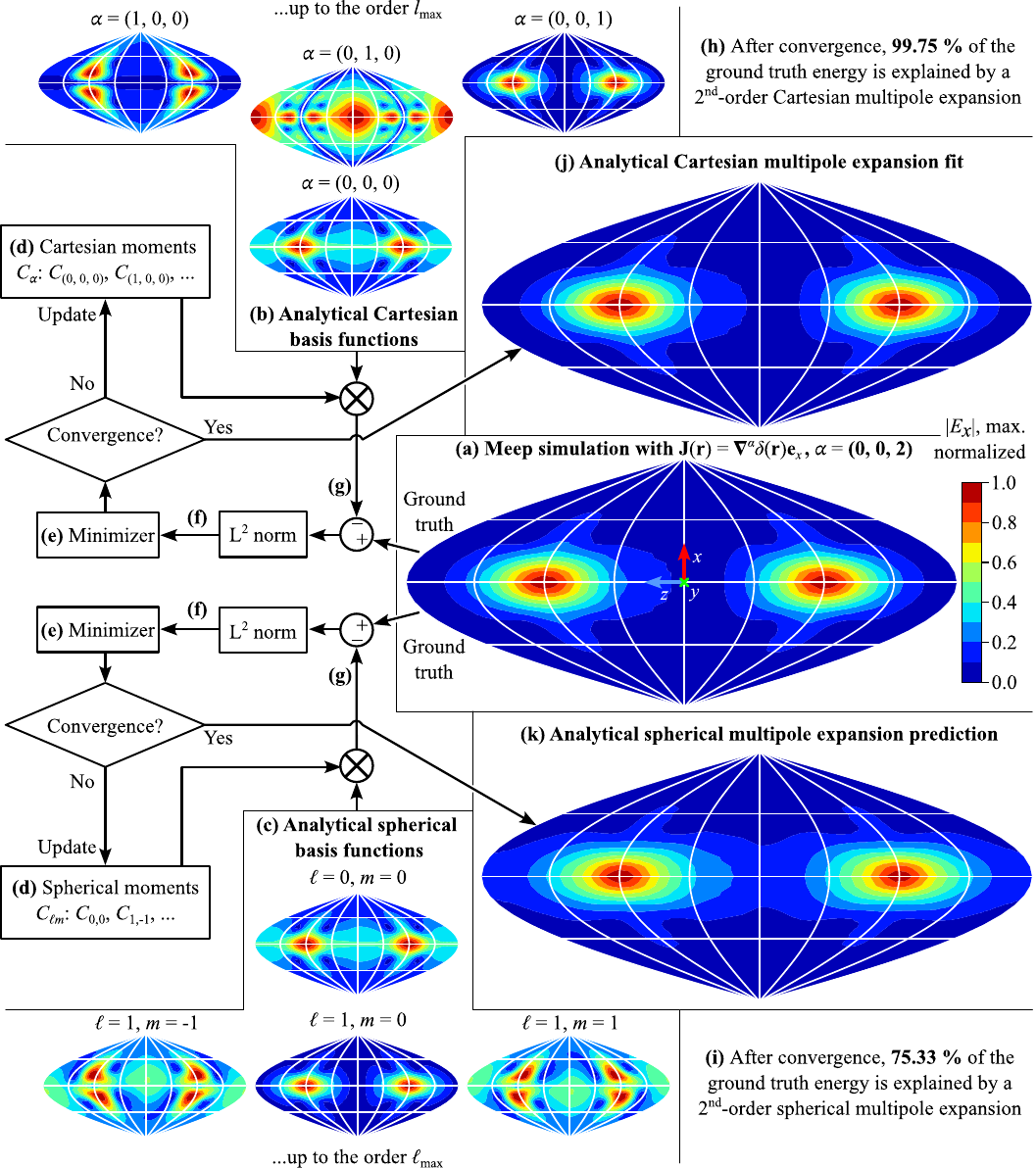}
\caption{\label{fig: Figure_B}}Flowchart of the inverse problem for $\varepsilon_{zz}=4$ and $\alpha=(0,0,2)$. 
    (a)~The ground truth data consists of all components (pictured: norm of the $x$ component, max-normalized; warm colors indicate high values) of the electric field radiated by the source $\mathbf{J}(\mathbf{r})=\nabla^\alpha \delta(\mathbf{r})\mathbf{e}_x$ simulated using Meep and sampled on a sphere surrounding the source. 
    The explaining data consists of the analytical Cartesian (b) and spherical (c) multipole expansion basis functions. These basis functions are multiplied by coefficients (d) determined by an optimization algorithm (e), whose goal is to minimize the $L^2$-norm of the error (f) between the ground truth and the multipole expansion (g). This error is normalized to the ground truth $L^2$-norm.
     Its complement after convergence represents the proportion of ground truth energy explained by the Cartesian (h) and spherical (i) multipole expansions. The fields obtained after convergence of the minimizer are shown in (j) for the Cartesian and (k) for the spherical multipole expansion.
\end{figure*}

The inverse problem consists in finding the spherical (resp. Cartesian) moments $C_{\ell m}$  (resp. $C_\alpha$) that minimize the squared $L^2$-norm of the error between the ground truth and the multipole expansion, normalized to the ground truth squared $L^2$-norm (energy). The minimization is carried out using the Nelder-Mead solver. After convergence, we evaluate the proportion of the ground truth energy explained by the multipole expansion. A schematic illustration of the problem is shown in Figure~(\ref{fig: Figure_B}).
\begin{figure}[t]
\includegraphics[width=.37\textwidth]{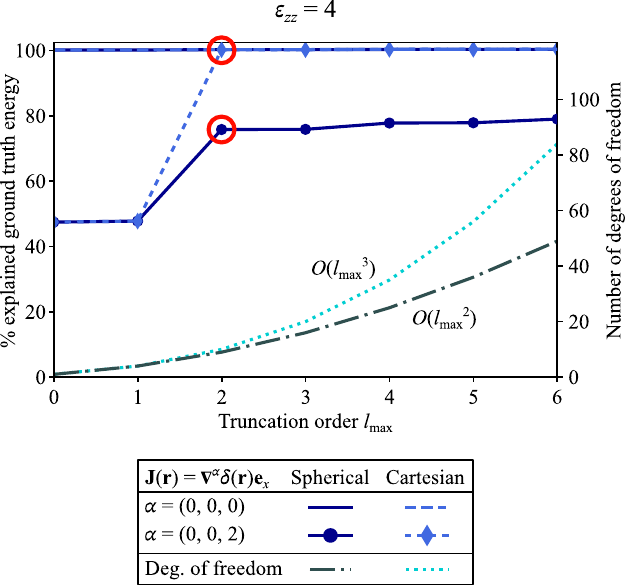}
\caption{\label{fig: Figure_A} Effect of the truncation order $\ell _\text{max}$ on the representation power of both multipole expansions (spherical: solid and solid-dotted; Cartesian: dashed) for both source types ($\alpha=(0,0,0)$: no markers; $\alpha=(0,0,2)$: markers). The number of degrees of freedom (i.e., the number of parameters determined by the inverse problem) is also given on the right-hand side axis. The results circled in red are illustrated in Figure~\ref{fig: Figure_B}.}
\end{figure}
\begin{figure}[t]
\includegraphics[width=.35\textwidth]{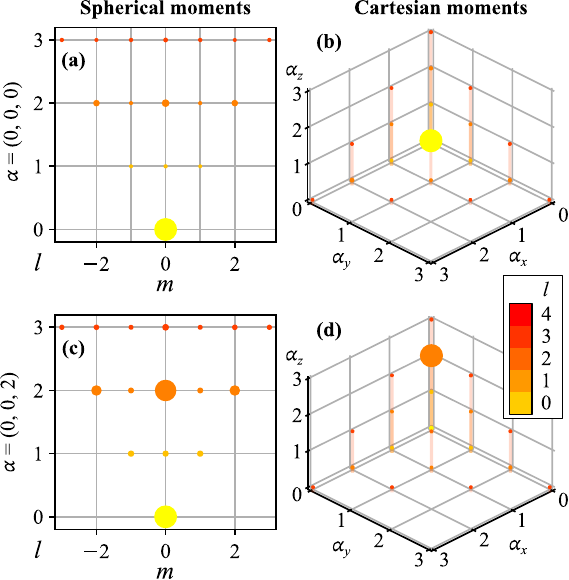}
\caption{\label{fig: Figure_C}Moments recovered by the inverse problem for $\varepsilon_{zz}=4$ and $\ell _\text{max}=4$. (a,~c) Spherical multipole expansions for (a) $\alpha=(0,0,0)$ and (c) $\alpha=(0,0,2)$. (b,~d) Cartesian multipole expansions for (b) $\alpha=(0,0,0)$ and (d) $\alpha=(0,0,2)$. The dot size is proportional to the maximum field amplitude of the corresponding moment. Warm colors indicate higher orders.}
\end{figure}
\begin{figure}[t]
\includegraphics[width=.35\textwidth]{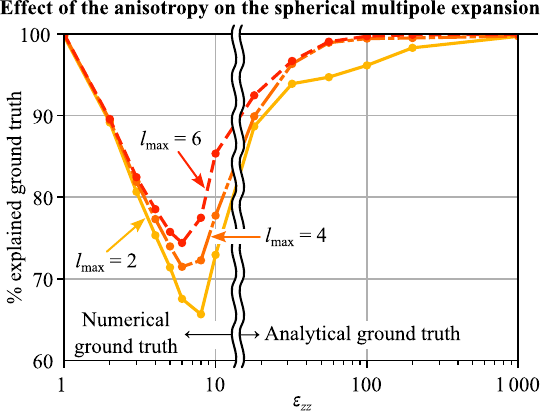}
\caption{\label{fig: Figure_D}Effect of the degree of anisotropy $\varepsilon_{zz}$ on the representation power of the spherical multipole expansion for the source $\mathbf{J}(\mathbf{r})=\nabla^{(0,0,2)}\delta(\mathbf{r})\mathbf{e}_x=\partial^2\delta(\mathbf{r})/\partial z^2\mathbf{e}_x$. Solid yellow line: $\ell_\text{max}=2$; solid-dashed orange line: $\ell_\text{max}=4$; dashed red line: $\ell_\text{max}=6$. For $\varepsilon_{zz}>10$, the analytical Green's function was used as ground truth data. }
\end{figure}

The results in Figure~(\ref{fig: Figure_A}) for an anisotropic medium with $\varepsilon_{zz}=4$ show that the dipole radiation (i.e., $\alpha=(0,0,0)$) is perfectly explained by both the spherical and Cartesian multipole expansions. This is expected, as the corresponding solutions $\textbf{f}_{0,0}$ and $\textbf{f}_{(0,0,0)}$ are proportional to each other and the monomial given by a constant is harmonic. However, the spherical multipole expansion fails to explain the non-harmonic octupole ($\alpha=(0,0,2)$) radiation, even with increased truncation orders $\ell _\text{max}$. Indeed, Figure~(\ref{fig: Figure_C}) shows that, while the original Cartesian multipole moment is accurately recovered by the inverse problem, the spherical multipole expansion needs four dominating moments ($(\ell ,m)=(0,0),(2,-2),(2,0),(2,2)$) to explain 75.4~\% of the ground truth energy. Moreover, Figure~(\ref{fig: Figure_D}) presents the effect of the degree of anisotropy on the representation power of the spherical multipole expansion. This figure shows that this lack of representation power worsens with increasing anisotropy up to $\varepsilon_{zz}=8$ for $\ell _\text{max}=2$. Indeed, 
in this regime, we observe the superposition of ordinary and extraordinary waves~\cite{felsen_radiation_1994} given by the dispersion relations
\begin{equation}
     \hspace{1.5em}k_0^2-\mathbf{k}^\top\mathbf{k}=0
\end{equation}
\begin{equation}
\varepsilon_{zz}k_0^2-\mathbf{k}^\top\overline{\overline{\varepsilon}}\mathbf{k}=0
\end{equation}
where $\mathbf{k}$ is the wavenumber vector, $\mathbf{k}^\top$ its transpose, and $k_0$ the free-space wavenumber. The second relation does not feature the Laplacian $- \mathbf{k}^\top\mathbf{k}$, but an anisotropic propagation $-\mathbf{k}^\top\overline{\overline{\varepsilon}}\mathbf{k}$ incompatible with the contraction in Equation~(\ref{eq: f alpha as lin comb of f lm}).

However, surprisingly, the representation power of the spherical multipole expansion for the non-harmonic octupole $\alpha=(0,0,2)$ increases with increasing anisotropy $\varepsilon_{zz}$ above $\varepsilon_{zz}=8$. Indeed, Figure~(\ref{fig: Figure_D}) shows asymptotes to a perfect representation for all orders $\ell _\text{max}$ as $\varepsilon_{zz}$ grows to $10^3$. The spherical expansion involves the moments $\ell =0,2$, $m=0$ to represent the fields in this large-anisotropy regime. 

To understand this regime, we start from the anisotropic vector wave equation for Green's function $\mathbf{G}(\mathbf{k})$ in wavenumber space, which reads (omitting the frequency dependence)
\begin{equation}
    \left(\mathbf{k}\mathbf{k}^\top -\mathbf{k}^\top\mathbf{k}\mathbf{I}+k_0^2\overline{\overline{\varepsilon}}\right)\mathbf{G}(\mathbf{k})=-\mathbf{I}
\end{equation}
with $\mathbf{I}$ the identity matrix. Solving for Green's function and taking the limit as $\varepsilon_{zz}\to\infty$, we obtain
\begin{equation}\label{eq: limit of G as eps -> inf}
    \mathbf{G}(\mathbf{k})\to \mathbf{G}_\infty(\mathbf{k})=-\underbrace{\frac{1}{k_0^2-k_z^2}\frac{1}{k_0^2-\mathbf{k}^\top\mathbf{k}}}_{g(\mathbf{k})}\left[(k_0^2-k_z^2)\mathbf{I}_{\perp}-\mathbf{k}_{\perp}\mathbf{k}_{\perp}^\top\right]
\end{equation}
where $\mathbf{I}_\perp=\mathbf{e}_x\mathbf{e}_x^\top+\mathbf{e}_y\mathbf{e}_y^\top$ and $\mathbf{k}_\perp=k_x\mathbf{e}_x+k_y\mathbf{e}_y$. 
As the propagation dynamics are determined by the denominator, we apply the contraction in Equation~(\ref{eq: laplacian of spherical sol}) to $g(\mathbf{k})$, obtaining
\begin{equation}\label{eq: def of h(k)}
    (k_0^2-\mathbf{k}^\top\mathbf{k})g(\mathbf{k}):=h(\mathbf{k})=\frac{1}{k_0^2-k_z^2}
\end{equation}
Multiplying by the right-hand side denominator and transforming back to spatial variables, the remainder function $h$ must satisfy
\begin{equation}
    k_0^2h(\mathbf{r})+\frac{\partial^2h(\mathbf{r})}{\partial z^2}=\delta(x,y)\delta(z)
\end{equation}
Solving this ordinary differential equation, we obtain (see Supplementary Note~2)
\begin{equation}\label{eq: h(r)}
    h(\mathbf{r})=c_+ \text{e}^{\text{i} k_0 z}+c_-\text{e}^{-\text{i} k_0 z}+\frac{1}{2k_0}\delta(x,y)\text{sign}(z)\sin(k_0 z)
\end{equation}
where $c_\pm$ are constants.
Thus, the contraction reads, by combining equations~(\ref{eq: limit of G as eps -> inf}) and~(\ref{eq: def of h(k)}) in the spatial domain,
\begin{equation}
    \nabla^2\mathbf{G}_\infty(\mathbf{r)}=-k_0^2\mathbf{G}_\infty(\mathbf{r)}-\left[\mathbf{I}_\perp\left(k_0^2+\frac{\partial^2}{\partial z^2}\right)+\nabla_\perp\nabla_\perp^\top\right]h(\mathbf{r})
\end{equation}
where $\nabla_\perp=\mathbf{e}_x\frac{\partial}{\partial x}+\mathbf{e}_y\frac{\partial}{\partial y}$. In other words, in the large anisotropy regime, we observe the same structure as in isotropic media. However, the singular region is no longer at the origin but is now defined by the function $h$. Arguably, the non-decaying traveling plane waves in $h$ can be discarded by setting $c_\pm=0$ to satisfy a radiation-like condition. In that case, the remaining support of $h$ shrinks to the $z$ axis because of the Dirac function $\delta(x,y)$. In turn, we find the anisotropic contraction outside this axis:
\begin{equation}
    \nabla^2\mathbf{G}_\infty(\mathbf{r)}=-k_0^2\mathbf{G}_\infty(\mathbf{r)}
\end{equation}
which explains the representation power of the spherical multipole expansion in the large anisotropic regime.

Finally, one might argue that it is always possible to represent fields on a sphere by a projection onto spherical harmonics.
While this is mathematically true, it is usually not helpful in physical problems. Indeed, the moments recovered by the projection do not correspond to those of the source responsible for the radiation. To illustrate, we project the $xx$ component of Green's function on the sphere described at the beginning of this section in isotropic ($\varepsilon_{zz}=1$) and anisotropic ($\varepsilon_{zz}=4$) media.
\begin{figure}[t]
\includegraphics[width=.5\textwidth]{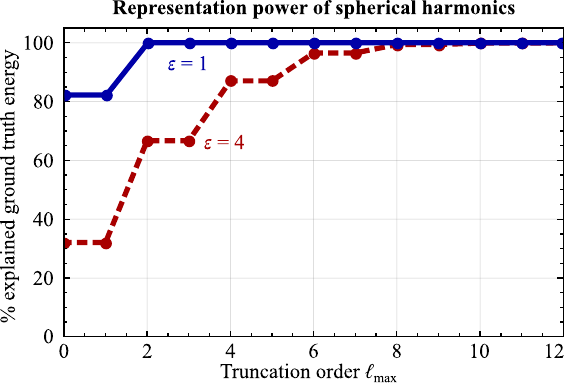}
\caption{\label{fig: Figure_E} Representation power of the spherical harmonics projected onto the $xx$ component of Green's function in isotropic ($\varepsilon_{zz}=1$, solid blue line) and anisotropic media ($\varepsilon_{zz}=4$, dashed red line).
}
\end{figure}
Figure~(\ref{fig: Figure_E}) shows the change in explained ground truth energy with increasing truncation orders $\ell _\text{max}$.  As expected, in isotropic media, a finite number of moments (up to second order) are needed to represent the field fully. However, moments of order up to eight are necessary in the considered anisotropic medium, even though the source is the same---only the field propagation changed.

It would also be possible to repeat this comparison by projecting the fields onto the restriction on the sphere of an arbitrary basis of polynomials of three variables. Since the spherical harmonics are formed by a restriction on the unit sphere of harmonic and homogeneous polynomials of three variables, this would consist of a Cartesian counterpart to the spherical harmonics. However, such a Cartesian projection would still be blind to the propagation anisotropy. Moreover, the added degrees of freedom would also be redundant, as it has been long shown that the spherical harmonics span all square-integrable functions on the sphere~\cite{axler_harmonic_2013}.
\section{Conclusion}
\label{seq: conclusion}
By analyzing the singular behavior of spherical and Cartesian multipole expansions, we applied harmonic function theory to find explicit links between both approaches in isotropic media while uncovering a lack of equivalence in anisotropic media. In those, the Cartesian approach is necessary, as illustrated in a radiation problem for electromagnetic waves. We showed that a spherical multipole expansion cannot represent the fields radiated by some sources, while a Cartesian multipole expansion can. Although a radical order increase improves the representation of the spherical expansion, the Cartesian expansion is still more efficient outside the large anisotropy regime.

Furthermore, by approaching the equivalence problem from the singular behavior of the solutions, we have shown how the Cartesian multipole expansion---obtained straightforwardly from Green's function---can be used to obtain the spherical one, which is not readily available in anisotropic media. This matters for cases where the dominant source feature is harmonic and, therefore, well-represented by the spherical approach. Also, the Laplacian-involving mapping from Cartesian to spherical approaches presented in Equation~(\ref{eq: f alpha as lin comb of f lm}) can be applied to many wave-like systems where the multipole expansion is already applied. Anytime the propagation is anisotropic, the Cartesian approach is expected to offer a better representation power and, ultimately, an improved understanding of the underlying physics. 

Note that the discrepancy between both approaches only appears for sources of order above one; this, together with the relatively recent rise of manufactured anisotropic media such as metamaterials, may explain the lack of prior research.
In addition to the electromagnetic application to optical and radio-frequency systems, our results are relevant to describing anisotropic interactions in plasma physics as discussed in the introduction. 
Similarly, the implications for cosmological studies on anisotropic baryon acoustic oscillations and the cosmic neutrino background 
 also highlight key examples where our findings can have an impact.


\section*{Acknowledgments}
E.L. thanks  Prof.~J.-Y.~Le~Boudec and Dr.~N.~Dietler for fruitful discussions. E.L. discloses support for the research of this work from the Technology Innovation Institute through agreement no. TII/DERC/2254/2021.

\section*{Author contributions}
Elias Le Boudec: conceptualization, methodology, software, validation, formal analysis, investigation, writing -- original draft, visualization; Toma Oregel-Chaumont: investigation, writing -- original draft, writing -- review \& editing; Farhad Rachidi: writing -- review \& editing, supervision, project administration, funding acquisition; Marcos Rubinstein: supervision; Felix Vega: supervision, project administration, funding acquisition.

\section*{Competing interests}
The authors declare no competing interests in relation to this work.

\section*{Data availability statement}
Source data for all figures are provided with the paper in the Supplementary Data~1.

\section*{Code availability statement}
The code that support the findings of this study is openly available: \cite{le_boudec_pynoza_2024} provides scripts to compute the equivalence between the spherical and Cartesian multipole expansions, and \cite{le_boudec_multipole_2024} provides the code used in the illustrating inverse problem.

\end{document}